\documentclass[prd,amssymb,twocolumn,nofootinbib,10pt]{revtex4-1}
\pdfoutput=1
\usepackage{amsmath,amsfonts,amssymb}
\usepackage{color,verbatim,graphicx,float}
\usepackage{bbold}

\begin{document}

\title{Quantum fluctuations of the compact phase space cosmology}

\author{Danilo Artigas${}^{*,\dagger}$}
\author{Sean Crowe${}^{*}$}
\author{Jakub Mielczarek${}^{*}$}
\affiliation{
${}^*$Institute of Theoretical Physics, Jagiellonian University, ul.\ {\L}ojasiewicza 11, 30-348 Krak\'{o}w, Poland \\
${}^\dagger$Institut d'Astrophysique Spatiale, UMR8617, CNRS, Universit\'e Paris-Sud, Universit\'e 
Paris-Saclay, b$\hat{a}$t. 121, 91405 Orsay, France
}

\begin{abstract}
In the recent article Phys.\ Rev.\ D {\bf 100}, no. 4, 043533 (2019) a compact phase space 
generalization of the flat de Sitter cosmology has been proposed. The main advantages of 
the compactification is that physical quantities are bounded, and the quantum theory is 
characterized by finite dimensional Hilbert space. Furthermore, by considering the 
$\mathbb{S}^2$ phase space, quantum description is constructed with the use $SU(2)$ 
representation theory. The purpose of this article is to apply effective methods to extract 
semi-classical regime of the quantum dynamics. The analysis is performed both without 
prior solving of the quantum constraint and by extracting physical Hamiltonian of the 
model. At the effective level, the results of the two procedures are shown to be 
equivalent. We find a nontrivial behavior of the fluctuations around the recollapse of the 
universe, which is distinct from what is found after quantization with the standard flat phase 
space. The behavior is reflected at the level of the modified Friedmann equation  
with quantum back-reaction effects, which is derived. Finally, an unexpected relation  
between the quantum fluctuations of the cosmological sector and the holographic Bousso 
bound is shown.
 
\end{abstract}  

\maketitle

\section{Introduction}

Twenty-three years after the surprising proposition by Einstein that spacetime might 
not be linear, Born suggested in 1938 that the conjugated momentum space might 
be curved as well \cite{Born49}. From then, many models with non-linear momentum 
space have been developed. As showed in the geometrical approach \cite{Ashtekar:1997ud}, 
many generalizations of quantum mechanics may arise at a more fundamental level 
by considering a non-linear phase space. In $q$-deformation theories or in non-commutative 
geometry, the non-linearity of momentum space is presented as a way to introduce 
non-commutativity of spacetime \citep{Snyder:1946qz, Majid:1999tc, Connes:1994yd}. 
On the other hand, such a geometry of momentum space may lead to a relative notion 
of locality \citep{AmelinoCamelia:2011bm, AmelinoCamelia:2011pe}. Beyond this non-linear 
property, a particularly interesting case is that of a compact momentum space. In Loop 
Quantum Gravity (LQG) \cite{Ashtekar:2004eh,Rovelli:1997yv} and Loop Quantum Cosmology 
(LQC) \cite{Bojowald:2008zzb}, compactification of the momentum space is introduced 
as a way to imply discreteness of length, area and volume operators, which \textit{de facto} 
resolves the Big Bang singularity \cite{Bojowald:2001xe}. This feature is the exact analogue 
of the discrete spectrum of momentum operator in usual quantum mechanics, when 
considering a compact position operator \cite{Rovelli:2014ssa}.

A natural and relevant way of extending those works could, therefore, be to consider 
both canonical variables to be non-linear, and eventually compact. Such studies have 
for example been conducted in the context of (3+1)-dimensional lattice Yang-Mills theory 
\citep{Riello:2017iti, Dittrich:2017nmq} and of LQG \cite{Haggard:2015ima}. The recently 
introduced Non-linear Field Space Theory (NFST) program \cite{Mielczarek:2016rax} aims 
to generalize and unify the phase-space compactification process to all field theories 
\citep{Bilski:2017gic, Mielczarek:2017ny}. In NFST, the phase space is compact and 
canonical variables are, therefore, bounded. This naturally satisfies Born's principle of 
finiteness \cite{Born:1934fy} and implies that, at the quantum level, the Hilbert space 
has finite dimension. In addition, a compact phase space opens a way to describe fields
with the use of spin variables \cite{Mielczarek:2016xql}.

One goal is to apply the NFST formalism to the gravitational field. This is an interesting 
application because it may allow to avoid infrared and ultraviolet singularities. For 
example, in \cite{Guimarey:2019lmn} it has been shown that the compactification of the 
phase space of the gravitational field at the level of a minisuperspace leads to a phase of 
recollapse, rather than a phase of infinite expansion, even in a universe dominated by 
dark energy (cosmological constant). Moreover, such a generalization is also motivated 
by loop quantum cosmology, where the phase space is already compactified in the 
momentum direction, leading to a cylindrical phase space. This momentum compactification 
implies to a resolution of the big bang singularity. However, in sufficiently flat universes 
it is expected that the universe will enter a phase of eternal expansion, which amounts 
to an infrared divergence. The compactification of the remainder of the phase space, 
cures this divergence, leading to a universe with many cycles of expansion and collapse.

The classical properties of a minisuperspace model with a compact phase space can 
be obtained in a straightforward manner, and in some situations exact solutions can be 
derived. After its quantization, it is possible to solve the model exactly in the case 
where the spin is small or when $s \sim \hbar$. Solutions for arbitrary spin $s$ are shown 
to exist but are, from a mathematical point of view, difficult to extract. However, in order 
to investigate the semiclassical limit, we need to solve the case where the spin is very 
large compared to $\hbar$, which is the goal of this paper.

Worth emphasizing is that the minisuperspace semiclassical sector, described 
in NFST cosmology by a state in $2s+1$ dimensional Hilbert space of a spin $s$, can 
be interpreted as a maximal spin subspace of a higher dimensional product space of 
fundamental representations of SU(2) - spin-1/2. This paves a way to both extract    
quantum cosmological sector from dynamics of elementary (inhomogeneous) spin-1/2 
degrees of freedom and to consider analog cosmological models employing condensed 
matter systems and quantum information processing technologies.

The objective of this article is to implement a semiclassical analysis of the model 
presented in \cite{Guimarey:2019lmn}. The Poisson bracket of this model is the 
bracket of $\mathrm{su}(2)$ algebra. Since this bracket is non-canonical the model 
is most easily quantized by the canonical quantization. However, the model is also 
constrained, owing to the fact that it is derived from general relativity (GR). The most 
systematic way of dealing with such a system at the effective level is by way of the 
canonical effective methods \cite{Bojowald:2009}. Therefore, we are formulating 
the dynamics of our system in terms of expectation values and their fluctuations, 
rather than in terms of wave functions. In the semiclassical limit, where higher 
order moments can be ignored this leads to a much more manageable system of 
ordinary differential equations rather than partial differential equations. 

In order to explain the idea of NFST cosmology, we first present a model 
of LQC in Sec. \ref{sec:LQC}, which can be considered as a 
limiting case of compact phase space cosmology. Instead of following the 
usual construction of LQC, we adopt the point of view of momentum 
polymerization which focuses on the geometry of the phase space. 
Generalizing such a construction, we present in Sec. \ref{sec:NFSC} the 
de Sitter model with compact phase space that has been derived and 
analyzed in details in \cite{Guimarey:2019lmn}. The core of this article 
is contained in Sec. \ref{sec:Semi}, where the semiclassical limit 
of our model is derived by means of canonical effective methods. 
Finally, in Sec. \ref{sec:Physics} we give physical analysis concerning 
such issues as: the number of elementary (inhomogeneous) degrees of 
freedom required to construct the semiclassical cosmological state, 
estimation of the magnitude of the quantum fluctuations  and the 
fate of the holographic Bousso bound are discussed. 
  
\section{de Sitter model in Loop Quantum Cosmology}
\label{sec:LQC}

\subsection{Classical Kinematics}

Let us first consider the case with the flat phase space. Since we are studying 
the canonical evolution of the Universe, we introduce a canonical physical co-volume 
$q$, depending on the scale factor $a$ and a fiducial volume of an elementary cell 
$V_0$, such that\footnote{The generalized coordinate $q\in \mathbb{R}$ 
is allowed to be negative as for taking into account different triad orientations.} 
$|q|:=V_0 a^3$. We introduce the corresponding canonical momentum variable 
$p$. In physical terms, this canonical momentum is proportional to the Hubble 
parameter. The phase space $(p,q)$ is therefore a symplectic manifold equipped 
with a closed 2-form $\omega$ which can be expressed in a Darboux form:
\begin{equation}
\omega = dp\wedge dq.
\label{DarbouxForm}
\end{equation}
The inversion of the symplectic form (\ref{DarbouxForm}) gives the usual Poisson bracket of the considered algebra:
\begin{equation}
\{f,g\} := (\omega^{-1})^{ij} \partial_i f \partial_j g = \frac{\partial f}{\partial q} \frac{\partial g}{\partial p} 
- \frac{\partial f}{\partial p} \frac{\partial g}{\partial q},
\label{ClassicalPoisson}
\end{equation}
for two arbitrary functions $f$ and $g$ defined on the phase space.

\subsection{Classical Dynamics}

In order to study the dynamics of our cosmological system, we now have to identify a 
suitable Hamiltonian $\mathcal{H}= NC$, such that any variable $f$ would evolve as
\begin{equation}
\dot{f}=\{f,\mathcal{H}\}
\label{dot}
\end{equation}
where $\dot{f}:=df/dt$ denotes the usual derivative along some time parameter $t$. 
In classical GR, the de Sitter Hamiltonian constraint for non-vanishing cosmological 
constant $\Lambda$ is
\begin{equation}
C_{\text{GR}}=q\left(-\frac{3}{4}\kappa p^2 +\frac{\Lambda}{\kappa} \right),
\label{ClassHam}
\end{equation}
where $q,p\in \mathbb{R}$, $\kappa:=8\pi G$ and $N$ is the lapse function.

Solving the dynamics of our system is equivalent to solving and integrating the 
constraint equation $\partial \mathcal{H}/\partial N=0$. This constraint is characteristic 
of any time-reparametrization invariant theory, such as GR. If we include some 
ordinary matter content, solving the constraint for the classical Hamiltonian 
(\ref{ClassHam}) leads to the well-known Big Bang singularity with cosmological 
constant $\Lambda$.

A common way for resolving the UV divergence (i.e. the Big Bang singularity) is that of loop quantum cosmology, which consists 
of the so-called momentum polymerization $p\in (-\infty,+\infty) \rightarrow \lambda p\in [-\pi,+\pi]$, where 
$\lambda$ is a physical length arising from discretization of lengths in loop quantum gravity \cite{Agullo:2013dla}, and we 
can generalize the Hamiltonian to be
\begin{equation}
C_{\text{LQC}}=q\left(-\frac{3}{4}\kappa \frac{\sin^2(\lambda p)}{\lambda^2} +\frac{\Lambda}{\kappa} \right).
\label{LQCHam}
\end{equation}

Notice that the polymerization of momentum $p\in [-\pi/\lambda, \pi/\lambda]$ usually performed in LQC 
doesn't change the symplectic structure. This is because one can always choose a local coordinate system 
on their symplectic manifolds to bring the symplectic structure into canonical form. In this example we have 
chosen to use a coordinate system with a canonical symplectic structure. However, after polymerization of 
both the generalized coordinate $q$ and momentum $p$ we will obtain a phase space which is equivalent 
to the phase space of spin, and the quantization of such a system is most easily done with a non-canonical 
Poisson structure. Therefore, in section (\ref{NPSC:Kin}) we will use a specific non-canonical bracket because 
it is more amenable to quantization.

As a way to characterize the rate of expansion of the Universe, it is convenient to introduce 
the Hubble factor $H:=\frac{\dot{a}}{a}=\frac{1}{3}\frac{\dot{q}}{q}$. By using (\ref{dot}), we 
can calculate the Hubble parameter in terms of the momentum. We can then insert this result 
into the Hamiltonian constraint $\partial \mathcal{H}_{\text{LQC}}/\partial N=0$, and derive the 
effective Friedmann equation for de Sitter universe in LQC \cite{Ashtekar:2006wn,Mielczarek:2008zv,Mielczarek:2010rq},
\begin{align}
H^2 =\frac{N^2\kappa}{3} \frac{\Lambda}{\kappa}\left(1-\frac{4\lambda^2}{3}\frac{\Lambda}{\kappa^2}\right) = \frac{N^2\kappa}{3} \rho_\Lambda \left(1-\frac{\rho_\Lambda}{\rho_c}\right),
\label{LQCFriedmann}
\end{align}
where we defined the cosmological constant energy density $\rho_\Lambda :=\Lambda/\kappa$ 
and the critical energy density $\rho_c := 3\kappa/4\lambda^2$. We notice that this model is equivalent 
to a classical de Sitter model with a renormalized cosmological constant. When taking ordinary matter 
into account, this effective Friedmann equation turns out to replace the Big Bang singularity by a 
Big Bounce instead, already at the semiclassical level \cite{Ashtekar:2006wn,Dzierzak:2009ip}.

\section{de Sitter model in Nonlinear Phase Space Cosmology}
\label{sec:NFSC}

\subsection{Kinematics}
\label{NPSC:Kin}
The NFST aims to generalize the fundamental field theories used in physics to the case of 
a compact phase space \citep{Mielczarek:2016rax, Mielczarek:2017ny}. The phase space 
$(q,p)$ is compactified into an ellipsoid, which is conveniently parametrized by spin vector 
$\vec{S}=(S_x,S_y,S_z)$, with components \cite{Mielczarek:2016xql}:
\begin{equation}
   \left \{
   \begin{array}{r c l}
S_x &=& S\cos\left(\frac{p}{R_1}\right)\cos\left(\frac{q}{R_2}\right)\\
S_y &=& S\sin\left(\frac{p}{R_1}\right)\cos\left(\frac{q}{R_2}\right)\\
S_z &=& -S\sin\left(\frac{q}{R_2}\right)
	\end{array}
	\right.
	\label{SpinVariable}
\end{equation}
with the angles $p/R_1 \in [-\pi,\pi]$ and $q/R_2 \in[-\pi/2, \pi/2]$. $S=\sqrt{\vec{S} \cdot \vec{S}}=R_1 R_2$ 
and $R_1$, $R_2$ are the two axes of the ellipsoid phase space. However, we can always make a canonical 
transformation which sets $R_1=R_2=\sqrt{S}$. Furthermore, it is convenient to keep these as two separate 
parameters, because it will allow us to study various limiting cases. For example, the limit $R_2\rightarrow \infty$ 
corresponds to the phase space of LQC. As shown in \cite{Guimarey:2019lmn}, the physical length $\lambda$ 
can be directly related to the radius of compactified momentum $\lambda:=1/R_1$, and the $R_{1,2}\rightarrow \infty$ 
limit is, therefore, equivalent to the classical affine phase space.

When considering such a spherical phase space, the symplectic 2-form should be replaced by
\begin{equation}
\omega = \cos{\left(\frac{q}{R_2}\right)} dp\wedge dq,
\end{equation}
and the Poisson bracket therefore reads
\begin{equation}
\{f,g\} := \frac{1}{\cos{\left(q/R_2\right)}} \left(\frac{\partial f}{\partial q} \frac{\partial g}{\partial p} 
- \frac{\partial f}{\partial p} \frac{\partial g}{\partial q}\right).
\label{NFSTPoisson}
\end{equation}
We, therefore, recover the standard kinematics in the limit when the phase space curvature goes to zero.

\subsection{Dynamics}

The compact phase space approach allows us to express the dynamics in terms of the spin variables 
(\ref{SpinVariable}). A relevant Hamiltonian constraint for describing the de Sitter model in the context 
of spherical phase space is the following one \cite{Guimarey:2019lmn}:
\begin{equation}
H_S=N C_S = N \frac{S_z}{R_1} \left[\frac{3}{4}\kappa \frac{S_y^2}{R_2^2}-\frac{\Lambda}{\kappa}\right],
\label{ConstraintCS}
\end{equation}
satisfying the following conditions:
\begin{equation}
   \left \{
   \begin{array}{l c l}
\lim_{R_2\rightarrow \infty} C_S &=& C_{\text{LQC}} \\
\lim_{R_{1,2}\rightarrow \infty} C_S &=& C_{\text{GR}}
	\end{array}
	\right. .
\end{equation}

For later convenience, we define the following constants:
\begin{equation}\label{delta}
\beta:= \frac{3}{2} \frac{\kappa}{R_1 R_2^2}, \quad \delta:=\frac{\rho_\Lambda}{\rho_c} =\frac{4\Lambda}{3\kappa^2 R_1^2},
\end{equation}
and fix the time gauge to $N=\frac{2}{\beta}$, such that constraint (\ref{ConstraintCS}) takes the convenient form:
\begin{equation}
C_S = S_z \left[ S_y^2 - \delta S^2 \right].
\label{NFSTHam}
\end{equation}
Employing the Hamilton equation for $q$, we deduce the Hubble factor at the kinematical level:
\begin{equation}
H = - \frac{1}{2} \frac{\kappa}{S^2 q} \frac{S_x S_y S_z}{\cos{(q/R_2)}}.
\label{NFSTHubbleKin}
\end{equation}
Imposing then the constraint $\partial H_S / \partial N =0$, the Friedmann equation can 
be derived \cite{Guimarey:2019lmn}:
\begin{equation}
H^2 = N^2 \frac{\Lambda}{3}\left(\frac{\sin(q/R_2)}{q/R_2} \right)^2 
\left[\frac{\cos^2\left(q/R_2\right)-\delta}{\cos^2(q/R_2)}\right].
\end{equation}

This generalized Friedmann equation ensures that the volume $q$ is bounded from above, 
implying a phase of recollapse for the Universe. In addition, it can be shown that in the limit 
$R_2 \rightarrow\infty$, we recover the Friedmann equations of LQC (\ref{LQCFriedmann}). 
We therefore suspect that, adding another matter field, the generalized Friedmann equation 
of NFST could lead to an oscillating Universe.

From the Hamiltonian (\ref{NFSTHam}) and the Poisson structure (\ref{NFSTPoisson}) we 
can determine the evolution of the spin variable:
\begin{equation}
   \left \{
   \begin{array}{r c l}
\dot{S_x}&=& 2 S_yS_z^2
+ \left(\delta S^2 -S_y^2\right) S_y, \\
\dot{S_y}&=& S_x \left( S_y^2-\delta S^2 \right), \\
\dot{S_z}&=&- 2 S_x S_y S_z.
	\end{array}
	\right.
\end{equation}

We can therefore fully describe the dynamics of our system by restricting the spin variable 
to the surface of constraint. This requires to solve the Hamiltonian constraint, here equivalent 
to $S_y=\pm\sqrt{\delta}S$, which then leads to the dynamical equations:
\begin{equation}
   \left \{
   \begin{array}{r c l}
\dot{S}_x&\approx& N\beta S_yS_z^2,\\
\dot{S}_y&\approx&0, \\
\dot{S}_z&\approx&- N\beta S_x S_y S_z.
	\end{array}
	\right.
\label{PhysSpin}
\end{equation}
where $N\beta=2$, but we here made the choice to write $N$ explicitly.

\subsection{Physical Hamiltonian}

We would now like to determine a physical Hamiltonian of our system $p_\phi$, that 
directly generates the dynamics of the spin variable on the surface of constraint (\ref{PhysSpin}). 
This physical Hamiltonian is by definition the generator of a physical time, that we will d
escribe as a scalar field $\phi$. It thus satisfies the new Poisson bracket relation 
$\phi':=\{\phi,p_\phi\}_{\phi,p_\phi}=1$, where the prime denotes derivative with respect 
to the physical time $(':=d/d\phi)$. By definition, under this time reparametrization $t\rightarrow\phi$, 
the lapse function becomes $N \rightarrow \tilde{N}= N (d\phi/dt)^{-1}$.

The equations of motion (\ref{PhysSpin}) are then given by 
\begin{equation}
\dot{S_i}=\frac{d\phi}{dt} S_i'= \frac{d\phi}{dt} \{S_i,p_\phi\}_{\phi,p_\phi}.
\end{equation}

We need to choose the specific physical time such that the resulting system of differential equations 
is actually a Hamiltonian system. To accomplish this we notice that $\dot{S}_{y}\approx 0$ implies 
the physical Hamiltonian can only depend on $S_y$. From the Leibniz rule then, it follows that the 
equations of motion with respect to the physical time should have the following behavior 
$S'_x\propto S_z$ and $S'_z\propto S_x$. To accomplish this we define our physical time implicitly 
in the following way:
\begin{equation}
\tilde{N} = N \left(\frac{d\phi}{dt}\right)^{-1} = N \left(S_z\right)^{-1}.
\end{equation}
The equations of motion (\ref{PhysSpin}) are then equivalent to:
\begin{equation}
   \left \{
   \begin{array}{r c l}
S_x'&=& \tilde{N}\beta S_yS_z^2 = 2 S_y S_z,\\
S_y'&=&0, \\
S_z'&=&- \tilde{N}\beta S_x S_y S_z = -2 S_x S_y.
	\end{array}
	\right.
\label{NewPhysSpin}
\end{equation}

This choice of physical time therefore makes the equations of motion much easier to deal with, 
furthermore this system of equations is now a Hamiltonian system, as required. In fact, not 
every choice of time parametrization will lead to equations of motion on the constraint surface 
that are Hamiltonian.  The choice we make here does, but we also make this choice because
it leads to a physical Hamiltonian that is quadratic and therefore amenable to quantization. 
However, one should notice that when the Universe will collapse, $\tilde{N}$ goes to infinity 
as the physical volume goes to 0 (i.e. $S_z\rightarrow 0$). In consequence, it will be more 
convenient to go back to the parameter time $t$ when analyzing the dynamics close to a 
singularity. From equations \eqref{NewPhysSpin}, one can easily deduce the corresponding 
physical Hamiltonian\footnote{Notice that the physical Hamiltonian is defined up to a constant, 
which we here set to $0$. That can always be done by redefining the origin of the physical 
time considered.} $p_\phi = S_y^2$.

Time in General Relativity is difficult to define, as it is a relational quantity \cite{Rovelli:2004tv,Bojowald:2010qpa}. 
The way to notice the flow of time is by describing how some objects evolve in relation with 
a distinguished degree of freedom that can play the role of time. In our simplistic model, the only 
matter present in the Universe is the cosmological constant. We've seen that the equations of 
motion \eqref{PhysSpin} satisfied the constraint $S_y^2=\delta S^2$. However, our derivation 
of a physical Hamiltonian therefore implies that $p_\phi=S_y^2= \delta S^2$. Choosing a time 
parametrization that makes our system Hamiltonian on the constraint surface therefore leads us 
to the interpretation that the cosmological constant is the momentum of a scalar field. This 
implied scalar field can then play the role of time.

\subsection{Quantization}

Since our model is expressed in terms of spin variables we can quantize it in the usual way: 
$S_i\rightarrow \hat{S}_i$. However, since we are working with a constrained system there 
are two ways we can proceed. We can quantize before or after gauge fixing. It turns out that 
these two approaches agree to first order quantum corrections. Quantizing after gauge fixing 
is most straightforward way to proceed. The deparametrized Hamiltonian corresponding to 
\eqref{NFSTHam} is: $p_{\phi}=S_y^2$, which has the simple quantization:
\begin{equation}\label{canham}
\hat{p}_{\phi}=\hat{S}_y^2.
\end{equation}

This system is actually solvable exactly. If we work in a basis that diagonalizes $S_y$. 
However, since we are only interested in semiclassical states, this analysis is not necessary for us.

We can instead choose to quantize the constraint \eqref{NFSTHam} before gauge fixing, we obtain 
the following quantum constraint:
\begin{equation}
\hat{C}_S = \left(\hat{S}_z \hat{S}_y^2\right)_{\text{Weyl}}-\left(\hat{S}_z \hat{p}_{\phi}\right)_{\text{Weyl}},
\label{QConstraint}
\end{equation}

where we made the choice to use Weyl ordering of all the operators, we have also made the choice 
to not include factorial powers of the $\hat{S}_i$ in the ordering choice. Moreover, $\left(...\right)_{\text{Weyl}}$ 
refers to the usual Weyl ordering, \textit{e.g.}:
\begin{equation}
\left(\hat{S}_z \hat{S}_y^2\right)_{\text{Weyl}}:=\frac{1}{3} \left(\hat{S}_z \hat{S}_y\hat{S}_y + \hat{S}_y \hat{S}_z \hat{S}_y + \hat{S}_y \hat{S}_y \hat{S}_z \right).
\end{equation}

We are not aware of a method to solve this constraint exactly for arbitrary choice of $s$. However, after 
fixing $s$ to a certain value, solving the constraint is equivalent to finding its null space, which can 
always be done, at least numerically for finite systems. As an example, exact solutions have been 
derived for low spin value $s=2$ in \cite{Guimarey:2019lmn}.

\section{Semiclassical limit of de Sitter in NFST}
\label{sec:Semi}

Solving the quantum constraint in full generality is a hard problem. Additionally, solving the 
deparametrized constraint is possible, however, we don't have a satisfactory way of selecting 
suitable initial conditions. Despite these difficulties, we are most interested in the behavior of 
our system in the semiclassical regime which can be accessed with the canonical effective 
methods. This is an interesting approach because it will allow us to see how the quantum 
fluctuations behave in the presence of a spherical phase space without having to solve the 
quantum dynamics exactly. 

\subsection{Semiclassical Mechanics}

After we have identified our classical phase space as the phase space of spin, our Poisson 
bracket takes the form:  $\left\{S_i,S_j\right\}=\varepsilon_{ijk}S_k$. The quantization of this 
bracket can be found with the usual canonical quantization. However, we aim to perform a 
semiclassical analysis of this system, which will involve studying the backreaction of the moments
onto the expectation values of the spin. We therefore need to understand the kinematics of the 
moments of the state. At the semiclassical level, the required Poisson brackets can be defined  
in a straightforward way \cite{Baytas:2017eq33}:
\begin{equation}
\left\{\langle \hat{A} \rangle,\langle \hat{B} \rangle\right\}=\frac{1}{i \hbar}\left\langle \left[\hat{A},\hat{B}\right]\right\rangle.
\end{equation}

Where this bracket is augmented with the Leibniz rule in order to satisfy all the properties of 
a Poisson bracket. Truncating our results at the leading semiclassical correction yields the following 
algebra of moments:
\begin{equation}\label{QuantBrack}
\begin{split}
\left\{S_i,S_j\right\}&=\varepsilon_{ijk}S_k,\\
\left\{\Delta\left(S_a S_b\right),S_c\right\}&=\varepsilon_{acd}\Delta(S_d S_b)+\varepsilon_{bcd}\Delta(S_d S_a),\\
\left\{\Delta(S_a S_b),\Delta(S_c S_d)\right\}&=\varepsilon_{ace}S_e \Delta(S_b S_d)+\varepsilon_{bce}S_e \Delta(S_a S_d)\\
&+\left(c \leftrightarrow d\right)+O(\hbar^{3/2}).
\end{split}
\end{equation}

Given a Hamiltonian, we can use this Poisson algebra to generate equations of motion on the 
semiclassical phase space. Moreover, this Poisson bracket is triply degenerate, admitting three 
Casimir functions. One of them takes the simple form: $\mathcal{C}_1=\langle \hat{S}^2 \rangle =\vec{S}^2+\Delta(\vec{S}^2)$. 
Due to this degeneracy, the expectation value of $\vec{S}^2$ is a constant of the motion, which 
is not surprising because this operator commutes with all Hamiltonians constructed from the spin 
operators. The second Casimir takes the form: $\mathcal{C}_2=2 S_i \Delta(S_i S_j)S_j+\Delta(S_i S_i)^2-\Delta(S_i S_j)\Delta(S_j S_i)$. 
The last and most interesting Casimir takes the form $\mathcal{C}_3=\det\left[\Delta(S_i S_j)\right]$. 
This implies that the volume of the wave packet in the spin space is conserved over time. This also 
gives us a crude measure of semiclassicality. As in the flat case a measure of semiclassicality is area 
of the wave packet in phase space. Here we can use the volume of the wave packet, which happens
to be a constant of the motion. Moreover, even at the level of kinematics we already have a divergence 
from the classical theory where the evolution is constrained to a sphere of radius $S$. Due to the 
quantum corrections this sphere actually becomes fuzzy and the classical spin is not conserved.
	
Given a Poisson bracket on the semiclassical phase space, we still need a Hamiltonian to generate 
dynamics. For consistency with the ordinary Schr\"odinger flow on the Hilbert space, the Hamiltonian 
on the semiclassical phase space takes the simple form $H_{Q}=\langle \hat{H} \rangle$, this expectation 
value can be expressed  as a sum of central moments of our quantum state, which can then be truncated 
at the leading semiclassical order. Furthermore, this Hamiltonian is  defined uniquely up to terms which
 depend only the Casimir function of the Poisson algebra.  From here the semiclassical dynamics can 
 be generated in the usual way: $\dot{K}=\left\{K,H_Q\right\}$, for some phase space function $K$.

\subsection{Effective constraints}

 In fact, setting the expectation value of \eqref{QConstraint} to zero is not the only constraint we need to 
 implement at the effective level. For leading order quantum corrections, we also need to implement to 
 following effective constraints: 
\begin{equation}
\begin{split}
C^{q}_s&=\langle \hat{C}_s \rangle=S_z(S_y^2-p_{\phi})+S_z \Delta(S_y^2)\\
&+2 S_y \Delta(S_z S_y)-\Delta(S_z p_{\phi}),\\
C_{S_i}&=\langle \hat{C}_s \hat{\delta} S_i\rangle_{W}=2 S_y S_z \Delta(S_y S_i)-S_z \Delta(p_{\phi}S_i),\\
C_{p\phi}&=\langle \hat{C}_s \hat{\delta} p_{\phi}\rangle_{W}=2 S_z S_y \Delta(S_y p_{\phi})-S_z \Delta(p_{\phi}^2),\\
C_{\phi}&=\langle \hat{C}_s \hat{\delta} \phi \rangle_{W}=2 S_z S_y \Delta(S_y \phi)-S_z \Delta(p_{\phi} \phi).
\end{split}
\end{equation}

A direct calculation shows that these constraints are preserved under evolution with respect to the coordinate time, 
that is, $\left\{C_s^q,C_i\right\}=0$. Moments involving $\phi$ can be chosen by a gauge choice and moments 
involving $p_{\phi}$ are selected by the constraints. In particular we have
\begin{equation}
\Delta(p_{\phi}S_z)=2 S_y \Delta(S_y S_z).
\end{equation}

Inserting this into the quantum constraint and solving for $p_{\phi}$, we find the following deparametrized Hamiltonian
\begin{equation}
p_{\phi}=S_y^2+\Delta(S_y^2)+O(\hbar^{2}).
\end{equation}
This is a quadratic Hamiltonian now, and furthermore, it is what we would expect from the quantization of deparametrized 
Hamiltonian \eqref{canham}. We also notice that both terms in this Hamiltonian are conserved separately. This can be seen 
by directly calculating the equations of motion with respect to the internal time:
\begin{equation}
\begin{split}
S'_y&=0,\\
\Delta(S_y^2)'&=0.
\end{split}
\end{equation}
The other equations of motion can be derived in the usual way: $K'=\left\{K,p_{\phi}\right\}$. Using the Poisson bracket 
\eqref{QuantBrack}  and the semiclassical Hamiltonian we find the following explicit equations of motion:
\begin{equation}
\begin{split}
S'_x&=2 S_y S_z+2 \Delta(S_y S_z),\\
S'_z&=-2 S_y S_x-2 \Delta(S_x S_y),\\
\Delta(S_x^2)'&=4 S_z \Delta(S_x S_y)+4 S_y \Delta(S_x S_z),\\
\Delta(S_z^2)'&=-4 S_x \Delta(S_y S_z)-4 S_y \Delta(S_x S_z),\\
\Delta(S_x S_y)'&=2 S_z \Delta(S_y^2)+2 S_y \Delta(S_y S_z),\\
\Delta(S_x S_z)'&=-2 S_y \Delta(S_x^2)+2 S_y \Delta(S_z^2),\\
&-2 S_x \Delta(S_x S_y)+2 S_z \Delta(S_y S_z),\\
\Delta(S_y S_z)'&=-2 S_y \Delta(S_x S_y)-2 S_x \Delta(S_y^2).
\end{split}
\label{EqMotion}
\end{equation}

\subsection{Analytic solutions for the quantum corrected spins}

The above system is quite complicated, and includes non-linear interactions between the classical 
and quantum parameters. However, due to the constant nature of $S_y$ and $\Delta(S_y^2)$ we 
are able to find analytic solutions for the whole system. We first notice that there is a subspace that 
obeys a linear set of equations of motion:
\begin{equation}
\begin{split}
S'_x&=2 S_y S_z+2 \Delta(S_y S_z),\\
S'_z&=-2 S_y S_x-2 \Delta(S_x S_y),\\
\Delta(S_x S_y)'&=2 S_z \Delta(S_y^2)+2 S_y \Delta(S_y S_z),\\
\Delta(S_y S_z)'&=-2 S_y \Delta(S_x S_y)-2 S_x \Delta(S_y^2).
\end{split}
\end{equation}

Since the coefficients of this system of differential equations are constant this system can be solved 
with standard methods, or with computer algebra. One caveat is that we have to Taylor expand our 
solutions to first order in the dimensionless parameter 
\begin{equation}
\epsilon: = \frac{\Delta(S_y^2)}{S_y^2}, 
\label{epsilon}
\end{equation} 
because this parameter is assumed to be small for semiclassical states. Rough estimates made below 
suggest it should be $O(\hbar/S)$, and we will ignore corrections that are $O(\epsilon^2)$. The general 
solution is available, however, too long to be printed here. We therefore present a solution corresponding to 
specific initial conditions. These initial conditions can be selected at the recollapse by assuming the 
state is evolving adiabatically at this time. We find the following solutions for the linear subspace,

\begin{widetext}
\begin{equation}\label{sols1}
\begin{split}
S_z(\phi)&=\sqrt{S^2-S_y^2}\Big[\cos{\left(2 S_y \phi\right)}+2 S_y \phi\left[ \sin{(2 S_y \phi)}-S_y \phi \cos{(2 S_y \phi)}\right]\frac{\Delta(S_y^2)}{S_y^2}\Big],\\
S_x(\phi)&=\sqrt{S^2-S_y^2}\Big[\sin{\left(2 S_y \phi\right)}-2 S_y \phi\left[\cos{(2 S_y \phi)}+S_y \phi \sin{(2 S_y \phi)}\right]\frac{\Delta(S_y^2)}{S_y^2}\Big],\\
\Delta(S_x S_y)&= \sqrt{S^2-S_y^2}\left[2 S_y \phi \cos{(2 S_y \phi)}-\sin{(S_y \phi)}\right]\frac{\Delta(S_y^2)}{S_y},\\
\Delta(S_z S_y)&=-\sqrt{S^2-S_y^2}\left[ \cos{(2 S_y \phi)}+2S_y \phi\sin{(2 S_y \phi)}\right]\frac{\Delta(S_y^2)}{S_y}.
\end{split}
\end{equation}
\end{widetext}

At this stage the semiclassical corrections to the classical motion have been derived. However, 
we would still like to solve for the remaining quantum moments. Inserting \eqref{sols1} in 
\eqref{EqMotion}, we find a system of linear homogeneous differential equations which can 
be solved and expanded using computer algebra to give solutions for the remaining moments. 
We fix the remaining initial conditions with our assumption that at the time of recollapse the 
moments are evolving adiabatically. Under these conditions the solutions read:

\begin{widetext}
\begin{equation}
\begin{split}
\Delta\left(S_x^2\right)&=\frac{1}{2}\Big[\left(S^2+S_y^2\right)+4 \left(S^2-S_y^2\right)S_y^2 \phi^2+\left(S^2-S_y^2\right)\Big[(-1+4 S_y^2 \phi^2)\cos{(4 S_y \phi)}-4 S_y \phi \sin{4 S_y \phi}\Big] \Big]\frac{\Delta\left(S_y^2\right)}{S_y^2}, \\
\Delta\left(S_x S_z\right)&=-\frac{1}{2}\left(S^2-S_y^2\right)\Big[4 S_y \phi \cos{(4 S_y \phi)}+\left(-1+4 S_y^2 \phi^2\right)\sin{(4 S_y \phi)}\Big]\frac{\Delta(S_y^2)}{S_y^2},\\
\Delta\left(S_z^2\right)&=\frac{1}{2}\Big[\left(S^2+S_y^2\right)+4 \left(S^2-S_y^2\right)S_y^2 \phi^2-\left(S^2-S_y^2\right)\Big[(-1+4 S_y^2 \phi^2)\cos{(4 S_y \phi)}-4 S_y \phi \sin{4 S_y \phi}\Big] \Big]\frac{\Delta\left(S_y^2\right)}{S_y^2}. 
\end{split}
\end{equation}
\end{widetext}

An interesting feature of these solutions is that the quantum corrected spins contain correction terms 
that are linear in the scalar field. This indicates a resonance between the classical and quantum degrees 
of freedom. Ordinarily this would signal a break down of perturbation theory. However, since we are 
only considering $\phi \in \left[-\frac{\pi}{4 S_y},\frac{\pi}{4 S_y}\right]$, this isn't an issue for this particular 
calculation. However, if we include ordinary matter in our calculations we will find a universe that goes 
through many cycles of bounces and contractions, allowing the effect from this resonance to accumulate 
over time.  However, we are working in a very simplified minisuperspace setting, so these resonances 
could be an artifact of this symmetry reduction. A more realistic model is then needed to verify whether 
or not this resonance is a robust feature.

For analyzing the behavior close to singularities it is most convenient to convert back to the proper time. 
This can be done by solving the differential equation: $\dot{\phi}=S_z$, this can be solved to 0th 
semiclassical order with the result: 
\begin{equation}
\phi= \frac{1}{2 S_y}\arcsin{\left(\tanh{\left(2 S_y \sqrt{S^2-S_y^2}t\right)}\right)}.
\end{equation}
We can then use this formula to convert our internal time results to the proper time.

\begin{figure}
\begin{center}
\includegraphics[scale=0.5]{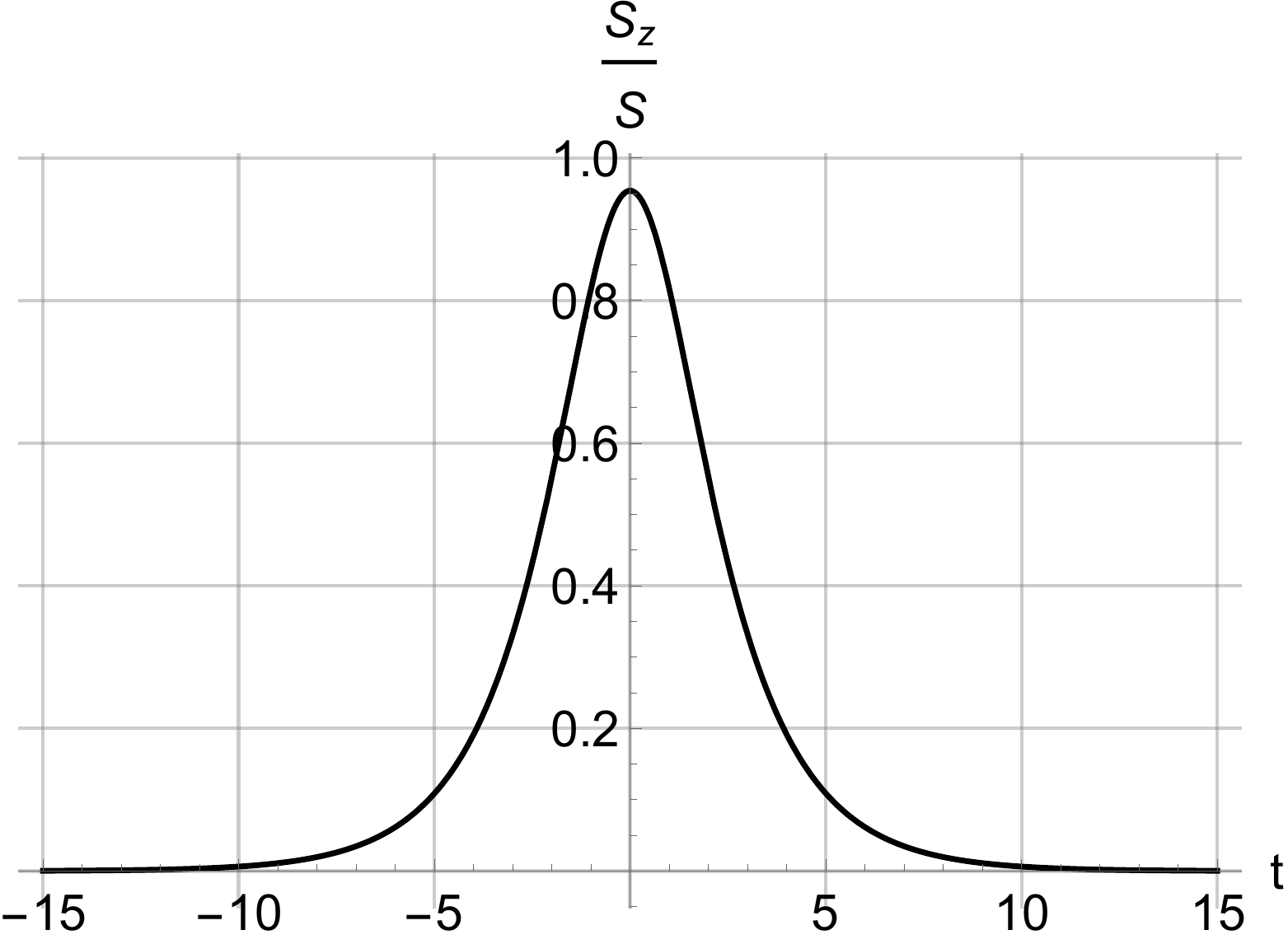}
\caption{The normalized z spin component as a function of the proper time. We see that when the magnitude 
of the proper time is much greater than one, then we have a universe which is either exponentially contracting 
or expanding. However, these two different solution branches are connected by a regime, where the phase 
space curvature effects dominate. In this plot we have chosen initial conditions such that $S_y/S=0.3$, and 
$\hbar/S=10^{-10}$. $\hbar$ being much less than $S$ puts us well into the perturbative regime and we can 
expect the numerical solutions for the classical coordinates to be close to the analytic solutions. Moreover, 
at the point of recollapse we have $S_x=0$ and $\Delta(S_y^2)=\frac{1}{2}\hbar S$. The adiabatic condition 
then fixes all other initial values. }
\end{center}
\end{figure}

\begin{figure}
\begin{center}
\includegraphics[scale=0.5]{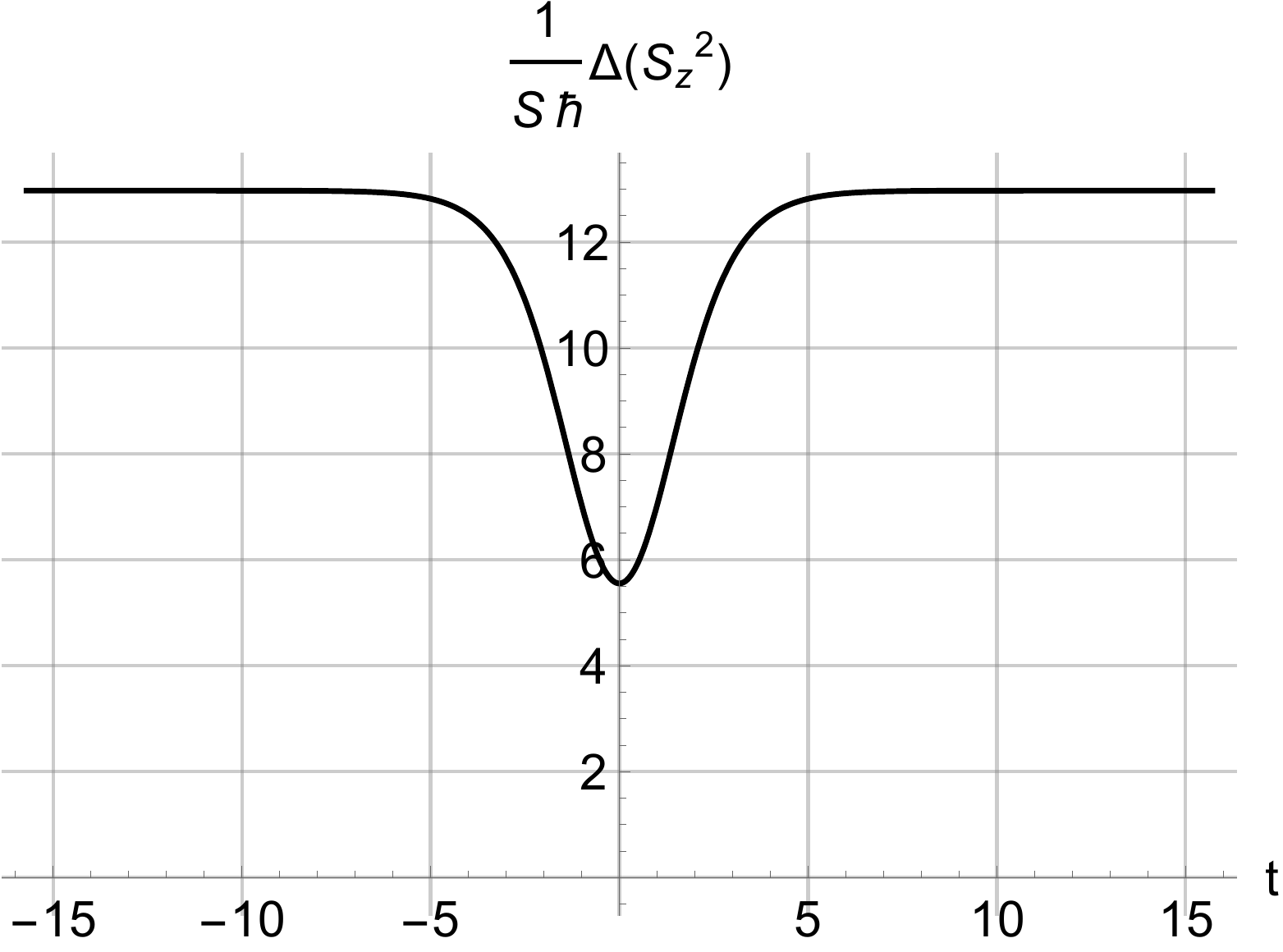}
\caption{Fluctuation of $S_z$ given our initial conditions as a function of the proper time. We notice a sharp 
decrease in the magnitude of the fluctuations close to the point of recollapse, contrary to what would be expected 
in a model with a flat phase space. Again, we are in the semiclassical regime because we have chosen 
$\hbar/S=10^{-10}$. This then implies that the we are in a regime where the quantum numbers are large 
because: $S^2 \sim s(s+1)\hbar^2$. This choice of initial conditions leads to moments that are parametrically 
small, and we can expect the corrections they give to be small. According to the discussion in (\ref{inicon}) 
we have chosen $\Delta(S_y^2)=\frac{1}{2}\hbar S$. After that the adiabatic condition fixes all other initial 
values at the point of recollapse.}
\label{fig:figure2}
\end{center}
\end{figure}

\begin{figure}
\begin{center}
\includegraphics[scale=0.5]{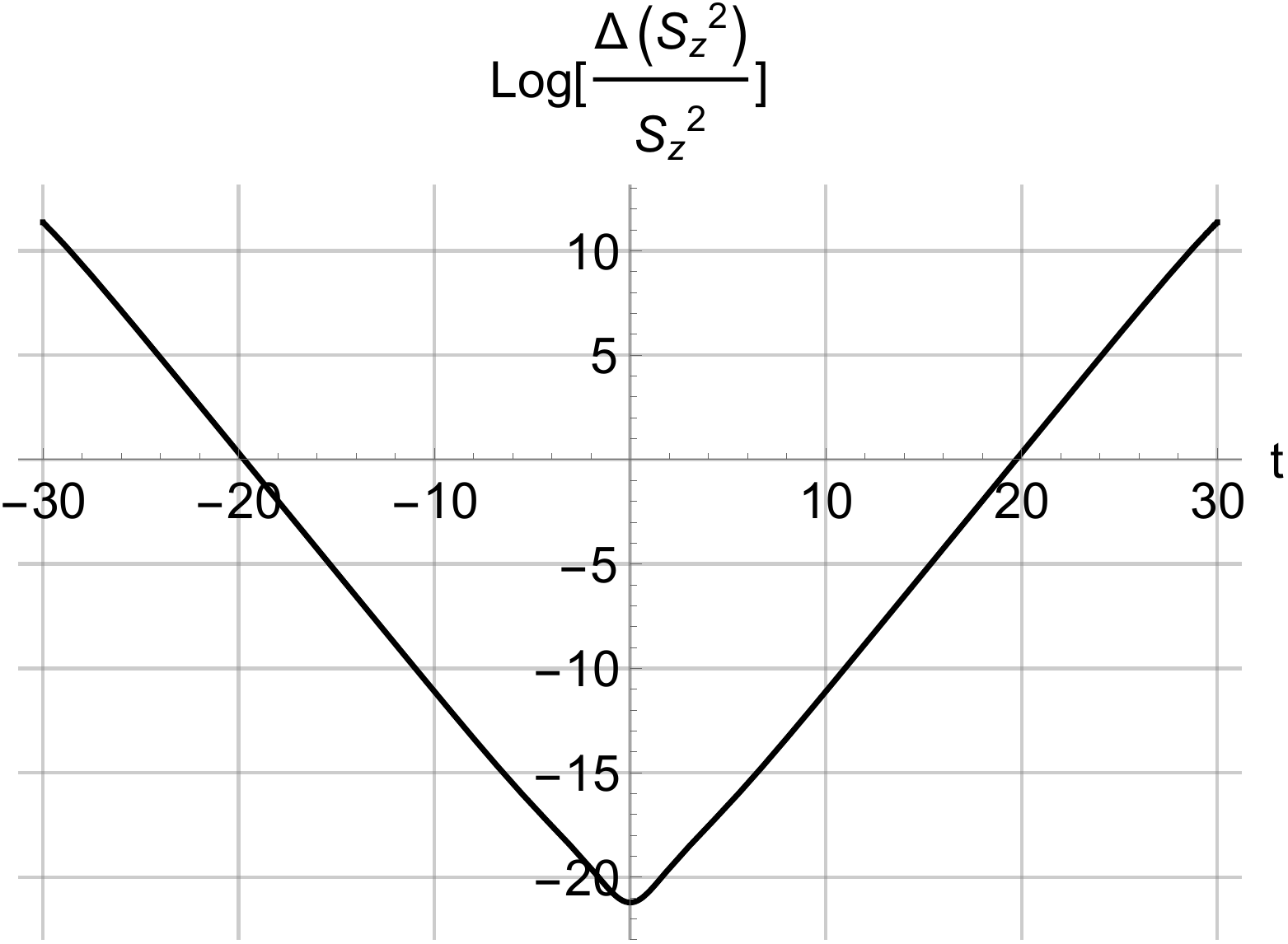}
\caption{Logarithm of the relative fluctuations relative to the proper time. The relative fluctuations are the 
smallest towards the point of recollapse, and grow exponentially as the universe shrinks in size. The 
quantum corrections therefore become dominant at a certain time in the early universe. For this plot we 
have chosen the same initial conditions as in Fig: \ref{fig:figure2}}
\end{center}
\end{figure}

\begin{figure}
\begin{center}
\includegraphics[scale=0.5]{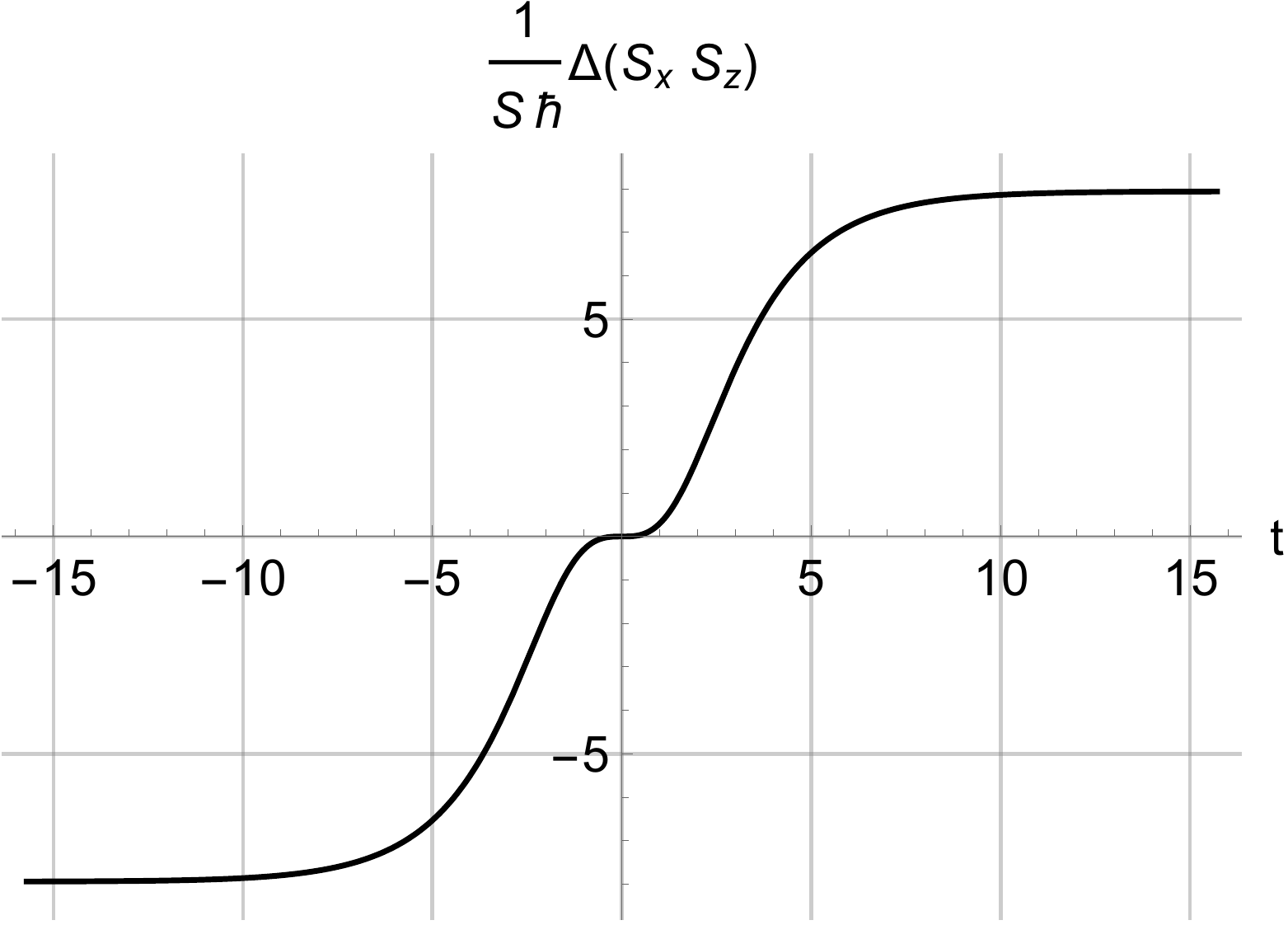}
\caption{Correlation function between $S_x$ and $S_z$ as a function of the proper time. Even though 
this correlation is zero at the point of recollapse for adiabatic initial conditions we see that nontrivial 
correlations can develop during the evolution. Again, we have chosen the same initial conditions as 
in Fig: \ref{fig:figure2}}
\end{center}
\end{figure}

\subsection{Effective Friedmann equations}

For phenomenological purposes it is important to derive a Friedmann equation which receives 
not only corrections from the phase space curvature, but also from the quantum corrections, 
which can have important phenomenological applications. We can use the equation of motion 
for $S_z$ to derive such an equation. Moreover, we can use our solutions for the moments and 
classical spins to express $S_z'$ in terms of only classical spins, or constants of motion. The 
equation of motion for $S_z$ is only corrected by the moment $\Delta(S_x S_y)$. Happily, this 
moment can be expressed in terms of the classical coordinates and the constants of motion:
\begin{equation}
\Delta(S_x S_y)=- S_x S_y \left(1- \frac{S_z}{S_x}\arctan{\left(\frac{S_x}{S_z}\right)}\right)\epsilon,
\end{equation}
where for the later convenience we have introduced parametrized relative quantum fluctuations 
of $S_y$ by the parameter $\epsilon$ defined in Eq. \ref{epsilon}.

We can now insert this solution into the equation of motion for $S_z$. We obtain:
\begin{equation}
S'_z=-2 S_y S_x\left[1-\left( 1- \frac{S_z}{S_x}\arctan{\left(\frac{S_x}{S_z}\right)}\right)\epsilon\right].
\end{equation}
Next, we perform a transformation back to the proper time, in order to make contact with the 
proper time Hubble factor. We obtain,
\begin{equation}
\dot{S}_z=-2 S_y S_x S_z \left[1-\left( 1- \frac{S_z}{S_x}\arctan{\left(\frac{S_x}{S_z}\right)}\right)\epsilon\right].
\end{equation}
We now undo the time reparametrization that we performed at the beginning of our analysis:
\begin{equation}
\dot{S}_z=-\frac{3}{2}\frac{\kappa S_y S_x S_z}{R_1 R_2^2}\left[1-\left( 1- \frac{S_z}{S_x}\arctan{\left(\frac{S_x}{S_z}\right)}\right)\epsilon\right]. 
\nonumber
\end{equation}
From the expression of $S_z$ \eqref{SpinVariable}, this implies that
\begin{equation}
\dot{q}=\frac{3}{2}\frac{\kappa}{S^2}\frac{S_x S_y S_z}{\cos{(q/R_2)}}\left[1-\left( 1- \frac{S_z}{S_x}\arctan{\left(\frac{S_x}{S_z}\right)}\right)\epsilon\right],
\nonumber
\end{equation}
agreeing with the kinematical Hubble factor \eqref{NFSTHubbleKin}, in the limit $\hbar\rightarrow 0$. 
Furthermore, we can express the Hubble parameters in terms of $q$ only:
\begin{align}
H^2&=N^2\frac{\Lambda}{3}\left(\frac{\sin{(q/R_2)}}{q/R_2}\right)^2\left[\frac{\cos{(q/R_2)^2}-\delta}{\cos{(q/R_2)^2}}\right] \nonumber \\
& \times \Big[1-2\epsilon \mathcal{F}(q)\Big]. \label{HubMod}
\end{align}
The pre-factor matches with what is found in the classical case, and the post factor contains 
quantum corrections, parametrized by the relative quantum fluctuations $\epsilon$. The function 
$\mathcal{F}(q)$ is given by:
\begin{equation}
\mathcal{F}(q)=1-\frac{\sin{(q/R_2)}\arctan{\left(\frac{\sqrt{\cos{^2(q/R_2)}-\delta}}{\sin{(q/R_2)}}\right)}}{\sqrt{\cos{^2(q/R_2)}-\delta}}.
\end{equation}
This function encodes the behavior of the semiclassical corrections. 

Taking the small phase space curvature limit we can make the approximation 
$\mathcal{F}\approx 1-\frac{\pi}{2 \sqrt{1-\delta}}\frac{q}{R_2}$. Furthermore, 
the entire Hubble parameter can be expressed in the small phase space curvature 
limit.
\begin{equation}
H^2=N^2 \frac{\Lambda}{3}\left[1-\delta-2\left[1-\frac{\pi}{2 \sqrt{1-\delta}}\frac{q}{R_2}\right]\epsilon \right],
\end{equation} 

The quantum effects therefore weaken the gravitational attraction at sufficiently small volumes, 
which is expected based on the expectation that quantum effects should make gravity repulsive 
at Planckian densities. As a possible phenomenological implication, we notice that in this limit, 
the quantum corrections contain a term that is proportional to the volume. The quantum corrected 
non linear phase space therefore predicts that there is a negative jerk, which can in principle be 
measured. However, to make an accurate prediction for this jerk, we would need to make our 
model more realistic by including other types of matter, as well as perturbations, going beyond 
the minisuperspace approximation.
 
\subsection{Cosmological NFST in the $\mathcal{C}_3=0$ case}
\label{inicon}

In order to justify some of our choices for initial conditions, and to gain some 
physical intuition for the physical evolution of our spin system, we consider 
the case where $\mathcal{C}_3=0$ such that we project our wavepacket on 
to a surface of constant spin, which is in turn equivalent to restricting our 
system to eigenstates of $\hat{S}^2$.

One way to implement this reduced system is to introduce a constraint to our 
semiclassical Hamiltonian. An alternative way, which gives more physical intuition, 
is to realize the spin algebra in terms of Casimir-Darboux  coordinates, and 
then to quantize the resulting canonical variables. Such a canonical parametrization is given by:
\begin{equation}
\begin{split}
S_x&=\sqrt{S^2-p_{\alpha}^2}\sin{\alpha},\\
S_y&=p_{\alpha},\\
S_z&=\sqrt{S^2-p_{\alpha}^2}\cos{\alpha}.
\end{split}
\end{equation}
Where we have $\left\{\alpha,p_{\alpha}\right\}=1$. It can be verified that this parametrization obeys the original bracket: $\left\{S_i,S_j\right\}=\epsilon_{ijk}S_k$.
By allowing $\alpha$ and $p_{\alpha}$ to fluctuate, we can relate fluctuations of the spin into fluctuations of these canonical coordinates.
\begin{equation}
\begin{split}
\hat{S}_i&=S_i+\delta \hat{S_i}\\
&=S_i\left[\alpha+\delta \hat{\alpha},p_{\alpha}+\delta \hat{p}_{\alpha}\right]\\
&\approx S_{i}\left[\alpha,p_{\alpha}\right]+\frac{\partial S_i}{\partial \alpha}\delta \hat{\alpha}+\frac{\partial S_i}{\partial p_{\alpha}}\delta \hat{p}_{\alpha}.
\end{split}
\end{equation}
This in turn implies:
\begin{widetext}
\begin{equation}\label{canonical_flucs}
\begin{split}
\Delta(S_y^2)&=\Delta(p_{\alpha}^2),\\
\Delta(S_x^2)&=\left(S^2-p_{\alpha}^2\right)\cos{^2(\alpha)}\Delta(\alpha^2)-\sin{(2 \alpha)}p_{\alpha}\Delta(\alpha p_{\alpha})+\frac{p_{\alpha}^2}{S^2-p_{\alpha}^2}\sin{^2(\alpha)} \Delta(p_{\alpha}^2),\\
\Delta(S_z^2)&=\left(S^2-p_{\alpha}^2\right)\sin{^2(\alpha)}\Delta(\alpha^2)+\sin{(2 \alpha)}p_{\alpha}\Delta(\alpha p_{\alpha})+\frac{p_{\alpha}^2}{S^2-p_{\alpha}^2}\cos{^2(\alpha)} \Delta(p_{\alpha}^2),\\
\Delta(S_y S_z)&=-\frac{p_{\alpha} \cos{(\alpha)}}{\sqrt{S^2-p_{\alpha}^2}}\Delta(p_{\alpha}^2)-\sqrt{S^2-p_{\alpha}^2}\sin{(\alpha)}\Delta(\alpha  p_{\alpha}),\\
\Delta(S_y S_x)&=-\frac{p_{\alpha} \sin{(\alpha)}}{\sqrt{S^2-p_{\alpha}^2}}\Delta(p_{\alpha}^2)+\sqrt{S^2-p_{\alpha}^2}\cos{(\alpha)}\Delta(\alpha p_{\alpha}),\\
\Delta(S_x S_z)&=-(S^2-p_{\alpha}^2)\cos{(\alpha)}\sin{(\alpha)}\Delta(\alpha
^2)-p_{\alpha}\cos{(2 \alpha)}\Delta(\alpha p_{\alpha})+\frac{p_{\alpha}^2}{S^2-p_{\alpha}^2}\cos{(\alpha)}\sin{(\alpha)}\Delta(p_{\alpha}^2).
\end{split}
\end{equation}
\end{widetext}
We then have the algebra:
\begin{equation}
\begin{split}
\left\{\alpha,p_{\alpha}\right\}&=1,\\
\left\{S^2,\bullet\right\}&=0,\\
\left\{\Delta(p_{\alpha}^2),\Delta(\alpha^2)\right\}&=-4\Delta(\alpha p_{\alpha}),\\
\left\{\Delta(p_{\alpha}^2),\Delta(\alpha p_{\alpha})\right\}&=-2\Delta(p_{\alpha}^2),\\
\left\{\Delta(\alpha^2),\Delta(\alpha p_{\alpha})\right\}&=2\Delta(\alpha^2).
\end{split}
\end{equation}
Finally, things are most simple if we use a faithful canonical parametrization of the moment algebra:
\begin{equation}
\begin{split}
\Delta(p_{\alpha}^2)&=p_{\sigma}^2+\frac{U}{\sigma^2},\\
\Delta(p_{\alpha} \alpha)&=\sigma p_{\sigma},\\
\Delta(\alpha^2)&=\sigma^2.
\end{split}
\end{equation}

The $U$ is the Casimir parameter of the algebra, which can be identified as the uncertainty 
of the canonical coordinates: $\Delta(p_{\alpha}^2)\Delta(\alpha^2)-\Delta(\alpha p_{\alpha})^2=U$. 
After this reduction of states, and the implementation of the canonical coordinate system we have 
the deparametrized Hamiltonian:

\begin{equation}\label{reducedham}
\begin{split}
H_Q=S_y^2+\Delta(S_y^2)
=p_{\alpha}^2+p_{\sigma}^2+\frac{U}{\sigma^2}.
\end{split}
\end{equation} 

Interpreting $\sqrt{U}$ as an angular momentum, we can identify this Hamiltonian as the Hamiltonian 
of a free particle in three dimensions written in cylindrical coordinates. Only the particle is not allow to 
get to close to the $\alpha$ axis, because the angular momentum is bounded from below by the 
Heisenberg uncertainty $\sqrt{U}\geq \hbar/2.$ This gives us some insight into the dynamics. For 
example, we expect $\sigma$ to grow without bound, and to generally only have one local minimum.

We can take things further by considering the eigenvalues and eigenvectors of the covariance 
matrix: $\Delta(S_i S_j)$. These quantities are important because the allow us to most easily 
vizualize the evolution of the ellipsoid in the phase space. The covariance matrix is three dimensional, 
so in the general case we might not expect to obtain nice analytical solutions. However, because 
of the reduction we made to the case $\mathcal{C}_3=0$, we expect one of the eigenvalues to be 
zero, because the wave packet is flattened on to the sphere of constant spin. Explicit calculations give:
\begin{widetext}
\begin{equation}\label{lambdas}
\begin{split}
\lambda_1&=0,\\
\lambda_2&=\frac{1}{2 (S^2-p_{\alpha}^2)\sigma^2}\Big[S^2 U+S^2 \sigma^2 p_{\sigma}^2+\left(S^2-p_{\alpha}^2\right)^2 \sigma^4\\
&+\sqrt{S^4 U^2+2 p_{\sigma}^2 S^4 U \sigma^2+S^2 \left(p_{\sigma}^4S^2-2(S^2-p_{\sigma}^2)^2U\right)\sigma^4+2 S^2 p_{\sigma}^2\left(S^2-p_{\alpha}^2\right)^2\sigma^6+\left(S^2-p_{\alpha}^2\right)^4 \sigma^8}\,\,\Big],\\
\lambda_3&=\frac{1}{2 (S^2-p_{\alpha}^2)\sigma^2}\Big[-S^2 U-S^2 \sigma^2 p_{\sigma}^2-\left(S^2-p_{\alpha}^2\right)^2 \sigma^4\\
&+\sqrt{S^4 U^2+2 p_{\sigma}^2 S^4 U \sigma^2+S^2 \left(p_{\sigma}^4S^2-2(S^2-p_{\sigma}^2)^2U\right)\sigma^4+2 S^2 p_{\sigma}^2\left(S^2-p_{\alpha}^2\right)^2\sigma^6+\left(S^2-p_{\alpha}^2\right)^4 \sigma^8}\,\,\Big].
\end{split}
\end{equation}
\end{widetext}

We remark that the only time dependent parameters in this expression are $\sigma$ and $p_{\sigma}$. 
Despite the complexity of these expressions, we have:
\begin{equation}\label{area}
\lambda_2 \lambda_3=U S^2.
\end{equation}
Indicating that the area of our wave packet on the sphere is conserved. Which is expected for semiclassical 
states evolving according to \eqref{reducedham}. Taylor expanding \eqref{lambdas} for large $\sigma$ (or 
equivalently the late time limit), we find:
\begin{equation}
\begin{split}
\lambda_2& \approx\frac{S^2}{S^2-p_{\alpha}^2}\left(p_{\sigma_o}^2+\frac{U}{\sigma_0^2}\right)+\left(S^2-p_{\alpha}^2\right)\sigma^2,\\
\lambda_3 &\approx \frac{S^2}{S^2-p_{\alpha}^2} \frac{U}{\sigma^2}.
\end{split}
\end{equation}
So generally the wave packet will be squeezed in one direction and stretched in the other, while 
remaining flat on the curved phase space, as one would expect from the case of a quantum particle 
evolving in flat two dimensional phase space.

Furthermore, this analysis allows us to obtain estimates for the initial conditions 
of the full quantum evolution. Starting with \eqref{area}, we notice that for symmetric 
states we have $\lambda^2=U S^2 \geq \frac{1}{4}\hbar^2 S^2$. Since, $\lambda$ is 
a good estimate for the size of the dispersions and because we will restrict ourselves 
to states that are not so extremely squeezed, we expect our initial moments to be of 
order $\hbar S$. It is therefore reasonable to take the initial condition: $\Delta(S_y^2)/(\delta S^2) 
=\epsilon \sim \hbar/(\delta S) $ in the full analysis where there is less analytic control. 
This epsilon parametrizes the quantum corrections in our model and therefore it is crucial 
that we make a good estimate for it. As noted below, $\epsilon$ is related to the relative 
fluctuations of the Hubble parameter and therefore we expect it to be very small.

In this reduced setting it is also instructive to consider the eigenvectors of the covariance 
matrix. Not surprisingly, the eigenvector with eigenvalue zero is $\nu_1=\vec{S}=
\left(\sqrt{S^2-p_{\alpha}^2}\sin{(\alpha)},p_{\alpha},\sqrt{S^2-p_{\alpha}^2}\cos{(\alpha)}\right)$. In the 
late time limit, we find the other eigenvectors have the approximate form: 
$\nu_2=\frac{\partial \nu_1}{\partial \alpha}$ and $\nu_3=\frac{\partial \nu_1}{\partial p_{\alpha}}$. 
Therefore, the wavepacket is spreading out in the direction of the motion in the phase space and 
contracting in the direction perpendicular to the motion. We find that these qualitative 
features are reproduced when the full dynamics are taken into account.

\section{Phenomenology and Bousso bound}
\label{sec:Physics}

The results which have been presented in the previous sections show that semiclassical 
homogeneous and isotropic gravitational sector can be constructed with the use of a spin 
degree of freedom. This has relevance from the point of view of extracting cosmological 
sector from a general (inhomogeneous) configuration of quantum gravitational system. 
Let us discuss this issue by assuming that elementary degrees of freedom of gravitational 
field are two dimensional quantum systems (qubits). This assumption is not crucial, and 
any higher dimensional elementary degrees of freedom can be considered, however, qubits 
are the smallest non-trivial quantum systems to which any higher dimensional quantum system 
can be always reduced to. The qubits are also corresponding to fundamental representations 
(spin-1/2) of the $SU(2)$ theory, which are utilized in such approaches to quantum gravity as 
LQG. 

Let us denote the Hilbert space of a qubit (spin-1/2) as $\mathcal{H}_{1/2}$, such that 
$\text{dim}\mathcal{H}_{1/2}=2$. Then, the Hilbert space of the system of $N$ qubits is 
\begin{equation}
\mathcal{H}_{\text{tot}}  = \bigotimes_{i=1}^N \mathcal{H}_{1/2}, 
\end{equation}
and has dimension  $\text{dim} \mathcal{H}_{\text{tot}}  = 2^N$. Without the lose of generality 
let us consider an even $N$. Then, the semiclassical homogenous and isotropic cosmological 
configuration is a state in a subspace $\mathcal{H}_{\text{cosm}} \subset \mathcal{H}_{\text{tot}}$ 
and is corresponding to the maximal spin. The maximal spin $s_{\text{max}}$, in the system of $N$ 
spins $1/2$ is $s_{\text{max}}=N/2$, such that    
\begin{equation}
\mathcal{H}_{\text{cosm}}=\mathcal{H}_{s_{\text{max}}},
\end{equation}
and $\text{dim} \mathcal{H}_{\text{cosm}} = 2s_{\text{max}}+1 = N+1$. Then, the total 
Hilbert space can be decomposed as follows $\mathcal{H}_{\text{tot}} = 
\mathcal{H}_{\text{cosm}}\otimes\mathcal{H}_{\text{inh}}$, where $\mathcal{H}_{\text{inh}}$ 
is the Hilbert space of the inhomogeneous sector having dimension:
\begin{equation}
\text{dim} \mathcal{H}_{\text{inh}} = 2^N-(N+1).
\end{equation}     

The spin $S$ present in our previous considerations can be now interpreted as $S=\hslash \sqrt{
s_{\text{max}}(s_{\text{max}}+1)} \approx  \hslash s_{\text{max}} = \frac{\hslash}{2} N$. So the
higher the number of degrees of freedom $N$ the bigger $S$ and in consequence, the spin 
description is closer to the standard case of GR.  

Based on the above considerations one can observe that, because cosmological sector is 
dependent on the parameter $S$, which in turn is a function of the total number of degrees 
of elementary freedom $N \sim S/\hslash$, analysis of the minisuperspace configuration 
allows to constrain the total number of degrees of freedom which are involved. Such analysis 
is especially interesting from the perspective of the constraint on the number of degrees of 
freedom in the Hubble volume implied by the \emph{holographic principle} \cite{Susskind:1994vu} 
and the holographic Bousso bound \cite{Bousso:1999xy,Bousso:2002ju}.

In the cosmological context, the holographic bound claims that the number of degrees of 
freedom in the Hubble volume $V_{\text{H}}$ - which we call $N_{\text{bulk}}$ - does not 
exceed the number of degrees of freedom at the boundary of the volume (which is the Hubble 
area $A_{\text{H}}$). By denoting the number of degrees of freedom at the Hubble area as $
N_{\text{boundary}}$, the Bousso bound can be stated as:
\begin{equation}
N_{\text{bulk}} \leq N_{\text{boundary}}.
\label{BoussoBound}
\end{equation}
While such inequality is satisfied, the bulk can be unambiguously described by the 
information at the boundary, realizing the holographic principe.

In order to answer if our considerations allow to tell something about the condition 
(\ref{BoussoBound}), let us consider the parameter (\ref{epsilon}), which appeared 
in the modified Friedmann equation (\ref{HubMod}). In the, semiclassical limit, 
far from the deep Planckian regime, where $p\ll R_1$ and  $q \sim R_2$ (so we are 
still far from the turning point), we have $S_y \approx \frac{S}{R_1} p $ (see Eq. 
\ref{SpinVariable}), which allows us to approximate Eq. \ref{epsilon} as 
\begin{equation}
\epsilon =  \frac{\Delta(S_y^2)}{S_y^2} \approx \frac{\Delta(p^2)}{p^2} \approx \frac{\Delta(H^2)}{H^2}.
\end{equation} 

The estimation of the relative fluctuations of the Hubble parameter is rather subtle. 
In what follows we will present an approach to the problem, which employs  
Planck scale fluctuations of the Hubble horizon

Namely, we can make an estimation of the relative fluctuations of the Hubble parameter, 
by changing variables to the Hubble radius $R_{\text{H}}:=1/H$, so that  
\begin{equation}
\begin{split}
\Delta(H^2)&:=\Big\langle \left(\hat{H}-H\right)^2 \Big\rangle
=\Big\langle \left(\frac{1}{\hat{R}_{\text{H}}}-\frac{1}{R_{\text{H}}}\right)^2 \Big\rangle\\
&=\Big\langle \left(\frac{1}{R_{\text{H}}+\widehat{\delta R_{\text{H}}}}-\frac{1}{R_{\text{H}}}\right)^2 \Big\rangle\\
&=\frac{1}{R_{\text{H}}^{4}}\Big\langle \left(\widehat{\delta R_{\text{H}}}\right)^2 \Big\rangle+O(\hbar^{{\color{black} 3/2}})\\
&\approx \frac{1}{R_H^{2}}\frac{\Delta(R_{\text{H}}^2)}{R_{\text{H}}^2}= H^2 \frac{\Delta(R_{\text{H}}^2)}{R^2_{\text{H}}}.
\end{split}
\end{equation}
This  implies that $\epsilon \approx \frac{\Delta(H^2)}{H^2} \approx \frac{\Delta(R_{\text{H}}^2)}{R^2_{\text{H}}}$. 
A reasonable assumption seem to be that quantum fluctuations of the Hubble horizon are tiny 
and of the order of the Planck length $l_{\text{Pl}} \approx 1.62 \cdot 10^{-35}$ m. Therefore, the standard deviation 
$\sigma (R_{\text{H}}) := \sqrt{\Delta(R_{\text{H}}^2)} \sim l_{\text{Pl}}$. Using this, as well as the fact that for de 
Sitter universe $R_{\text{H}} = \sqrt{\frac{3}{\Lambda}}$ (which is approximately satisfied in the case considered here), 
we can estimate that:
\begin{equation}
\epsilon \approx \frac{\Delta(S_y^2)}{\delta S^2} \approx \frac{\Delta(R_{\text{H}}^2)}{R^2_{\text{H}}} \sim l^2_{\text{Pl}} \Lambda,
\label{epsilonlambda}
\end{equation} 
where in the first approximation the equation of constraint $S_y^2\approx \delta S^2$ has been used. Taking the 
current values of the Hubble factor $H_0=67.8 \pm 0.9 \frac{\text{km}}{\text{s}\cdot\text{Mpc}}$ \cite{Ade:2015xua}  and the 
parameter $\Omega_{\Lambda} = \frac{\Lambda}{3 H_0^2} =0.69 \pm 0.01$ \cite{Ade:2015xua} we can estimate that: 
\begin{equation}
\epsilon \sim 10^{-122}.  
\label{epsiloncurrent}
\end{equation} 

Assuming that the momentum compactification scale (related to $R_1$) is Planckian, as one might expect from results in LQG,
implies that $\rho_c \sim \rho_{\text{Pl}}:= m^4_{\text{Pl}}$. In consequence, 
\begin{equation}
\delta:= \frac{\rho_{\Lambda}}{\rho_c} \sim \frac{\rho_{\Lambda}}{\rho_{\text{Pl}}} \sim l^2_{\text{Pl}} \Lambda  \sim \epsilon. 
\label{deltaepsilon}
\end{equation}

Furthermore, from Sec. \ref{inicon} we have the estimate, $\Delta{S_y^2}=\frac{1}{2}\hbar S$, which together with Eq. 
\ref{deltaepsilon}, allows us to rewrite Eq. \ref{epsilonlambda} into 
\begin{equation}
\frac{S}{\hbar}\sim  \frac{1}{\epsilon^{2}} \sim 10^{224},
\end{equation}
where the numerical value corresponds to the current estimate, employing Eq. \ref{epsiloncurrent}. As discussed 
earlier in this section, the value of $S$ can give an estimate for the number of degrees of freedom in the bulk, i.e. 
$ S \sim N_{\text{bulk}} \hslash$. Based on this we find,  $N_{\text{bulk}} \sim 1/\epsilon^{2} \sim 10^{224}$.

On the other hand, the number of degrees of freedom at the boundary (Hubble horizon) can be estimated by 
dividing the area of the Hubble horizon, by the Planck cells of the area $l^2_{\text{Pl}}$ \cite{Artymowski:2018pyg}:
\begin{equation}
N_{\text{boundary}} \approx \frac{A_{\text{H}}}{l^2_{\text{Pl}}} =\frac{4\pi R^2_{\text{H}}}{l^2_{\text{Pl}}} \sim 
\frac{1}{l^2_{\text{Pl}} \Lambda} \sim \frac{1}{\epsilon} \sim 10^{122}.
\end{equation}

Therefore, in the considered case:
\begin{equation}
N_{\text{bulk}} \sim N_{\text{boundary}}^2, 
\end{equation}
and the  number of degrees of freedom in the bulk is much greater than those on the boundary, violating the holographic 
Bousso bound (\ref{BoussoBound}).

In order to satisfy the Bousso bound we might try to generalize our estimate of $\Delta(S_y^2)$. We only have two 
parameters with dimension of action, so we can consider the power-law ansatz: $\Delta(S_y^2)\sim\hbar^2 (S/\hslash)^{k}$, 
where $k\leq 2$. The upper bound on the values of $k$ comes from the fact that the relative fluctuations 
$\frac{\Delta(S_y^2)}{S^2}  \sim (S/\hslash)^{k-2}$, and the values $k>2$ imply lack of semiclassicality of the state. 
For $k=1$ the previously case, violating the Bousso bound, is recovered. Another spacial case is $k=0$, which implies 
$\Delta(S_y^2)\sim \hbar^2$ and corresponds to a strongly squeezed state. For $k<0$, while the relative fluctuations
of $S_y$ decrease with the increase of $S$ (guaranteeing semiclassicality), the squeezing of state is very large, which 
has to be balanced by a large fluctuations of the other spin components.  Therefore, while the despite the $k<0$ 
satisfies the Bousso bound, since we have 
\begin{equation}
N_{\text{bulk}} \sim N_{\text{boundary}}^{\frac{2}{2-k}}, 
\end{equation} 
with the power $\frac{2}{2-k}<1$, such case is rather disfavored on the physical ground. The remaining case is $k=0$, for which 
\begin{equation}
N_{\text{bulk}} \sim N_{\text{boundary}}, 
\end{equation} 
and the Bousso bound has chance to be satisfied by a proper choice of the proportionality factor. 

Concerning  the issue of semiclassicality for the $k=0$ case, let us resort to our analysis of the quantum 
fluctuations of the undeformed spherical phase space, presented in Sec. \ref{sec:Semi}E. This is a reasonable 
case, especially if we assume that the state is condensing in the $S=s_{\text{max}} \hbar$ sector of the Hilbert space. 
The squeezing with $k=0$ implies $\Delta(p_{\alpha}^2)\sim \hbar^2$, but then the Heisenberg uncertainty 
principle for canonical variables implies: $\Delta(\alpha^2)\sim 1$. Given our canonical mapping this implies: 
$\frac{\Delta(S_x^2)}{S^2}\sim \frac{\Delta(S_z^2)}{S^2}\sim 1$. The character of such a state which is squeezed 
and satisfies the Bousso bound, is therefore not semiclassical.

\section{Summary}

In summary, our semiclassical analysis of a de Sitter model with a compact ($\mathbb{S}^2$) 
phase space has given us insight into the behavior of the quantum fluctuations in the model. 
In particular, we were able to find analytic solutions for the leading order moment 
corrections, despite the non-linearity of the effective system. These analytic solutions 
agree well with standard results of quantum fluctuations in a de-Sitter model with a 
flat phase space, and make new predictions in the regime where the classical coordinates 
are the same order as the phase space radius of curvature. While qualitative, these predictions 
are interesting, and possibly indicate future research directions.

Unfortunately, it is hard to make concrete quantitative phenomenological predictions 
in this setting, due to the highly simplified nature of the model. However, our solutions 
for the semiclassical corrections do indicate several interesting qualitative predictions. 
For example, our solutions indicate that the classical variables would not be periodic 
in a cyclic universe, and in particular,  these classical variables are resonating with the 
quantum corrections. The time constant for this resonance, is several lifetimes of the 
universe. Within this simplified setting of this model, this indicates the semiclassical 
states are not a stable variational class. A possible way around this issue would be to 
introduce different types of ordinary matter, or to include density perturbations. However, 
this would complicate the model, possibly making analytic solutions impossible to find.

Furthermore, our solutions have several qualitative features which could have 
phenomenological implications. Our calculation of the NFST Hubble parameter in the 
presence of quantum corrections indicates that there is a negative jerk which
could become relevant before one would expect based on the classical NFST. 
Moreover, for the initial conditions under consideration, we observe a sharp decrease 
in the quantum fluctuations close to the point of recollapse. 

Finally, the obtained results are discussed from the perspective of the 
number of degrees of freedom involved in construction of the semiclassical cosmological 
state and the fate of the holographic principle. We find that the semiclassical state 
supported by the considerations presented in this article violates the Bousso bound. 
However, a state which may satisfy the holographic bound is still acceptable based 
on physical considerations. This opens an intriguing possibility to build a quantum 
(compact phase space) version of de Sitter model satisfying the Bousso bound.

\section*{Acknowledgements}

Authors are supported by the Sonata Bis Grant No. DEC-2017/26/E/ST2/00763 of the National Science Centre Poland.

\end{document}